\documentclass[a4paper]{article}

\begin{document}

\title{\textbf{The Relativistic Quantum Law of motion for a Particle with Spin $1/2$. }}

\author{T.~Djama\thanks{E-mail:
{\tt djam\_touf@yahoo.fr}}}

\date {November 4, 2003}

\maketitle

\centerline {$14$, rue Si El Hou{\`e}s, B{\'e}ja{\"\i}a $06000$,
Algeria}

\begin{abstract}
\noindent In this paper, we introduce a deterministic approach of
quantum mechanics for particles with spin ${1 \over 2}$ moving in
one dimension. We present a Lagrangian of a spinning particle ($s=
{1 \over 2}$), and deduce the expression of the conjugate momentum
related to the velocity of the particle.
\end{abstract}

\vskip\baselineskip



\newpage

\vskip\baselineskip \noindent \textbf{1- Introduction}
\vskip\baselineskip
%
%

Recently, in the frame of a deterministic approach of quantum
mechanics, we derived the two relativistic Quantum Stationary
Hamilton-Jacobi Equation for a particle with spin ${1 \over 2}$
\cite {Djama1},
\begin{eqnarray}
{1 \over 2m_0}\left({dS_0 \over dx }\right)^2- {\hbar^2 \over
4m_0} \{S_{0},x\}+ {\hbar^2 \over 2m_{0}}(E-V+m_{0}c^2)^{1 \over
2}\ .
\hskip25mm&& \nonumber\\
\ .{d^2 \over dx^2} \left[(E-V+m_{0}c^2)^{-{1 \over 2}} \right]+
{1 \over 2m_0c^2}\left[m_0^2c^4 -(E-V)^2\right]=0\; ,
\end{eqnarray}
and
\begin{eqnarray}
{1 \over 2m_0}\left({dZ_0 \over dx }\right)^2- {\hbar^2 \over
4m_0} \{Z_{0},x\}+ {\hbar^2 \over 2m_{0}}(E-V-m_{0}c^2)^{1 \over
2}\ .
\hskip25mm&& \nonumber\\
\ .{d^2 \over dx^2} \left[(E-V-m_{0}c^2)^{-{1 \over 2}}\right] +
{1 \over 2m_0c^2}\left[m_0^2c^4 -(E-V)^2\right]=0\; ,
\end{eqnarray}
where
$$
\{f(x),x\}=\left[{3 \over 2} \left({df \over dx }\right)^{-2}
\left({d^2 f \over dx^2}\right)^2-\left({df \over dx}\right)^{-1}
\left({d^3 f \over dx^3}\right) \right]
$$
represent the schwarzian derivative of $f(x)$ with respect to $x$.
Eqs. (1) and (2) represent the two Relativistic Quantum
Stationary Hamilton Jacobi Equations for Spinning particle $(s={1
\over 2})$ (QSHJES$_{1 \over 2}$. One of these equations
correspond to the projection $m_{s}=+{1 \over 2}$ of the spin,
when the other correspond to the projection $m_{s}=-{1 \over 2}$.
It follows that the reduced actions $S_{0}$ and $Z_{0}$ correspond
to the two projections of the spin.

\noindent To establish these equations, we started from the Dirac
Spinors Equation written in one dimension in Ref. \cite {Kudo} as
\begin{equation}
-i\hbar c \ \ \sigma_{x} \ \ {d \psi \over dx} \ =
(E-V(x)-\sigma_{z}\ m_{0}c^2) \ \psi
\end{equation}
where
\begin{equation}
\sigma_{x}=\begin{pmatrix}
      { 0&1  \cr
        1&0  \cr}
 \end{pmatrix}; \ \
 \sigma_{y}=\begin{pmatrix}
      { 0&-i  \cr
        i&\ 0  \cr}
 \end{pmatrix}; \ \
 \sigma_{z}=\begin{pmatrix}
      { 1&\ 0  \cr
        0&-1  \cr}
 \end{pmatrix}
\end{equation}
are the Pauli matrix. $\psi = \begin{pmatrix}
      { \theta  \cr
        \phi  \cr}
 \end{pmatrix}$ is a the matrix of the wave functions $\theta$ and
 $\phi$.

\noindent In Ref. \cite {Djama1}, we established the solutions of
Eqs. (1) and (2). We written the reduced actions as
\begin{equation}
S_0=\hbar \arctan\left(a\ {\theta_{1}(x) \over
\theta_{2}(x)}+b\right) \; ,
\end{equation}
\begin{equation}
Z_0=\hbar \arctan\left(d\ {\phi_{1}(x) \over \phi_{2}(x)}+e\right)
\; ,
\end{equation}
where $a, b, d$ and $e$ are real constants.
($\theta_{1}$,$\phi_{1}$) and ($\theta_{2}$, $\phi_{2}$) are two
real and independent solutions sets of the Dirac Spinors Equation
(Eq. (3)) \cite {Djama1}.

\noindent All these results make it possible to introduce a
dynamic formulation of relativistic quantum mechanics as it is
done in Refs. \cite {B-D1,B-D2,B-D3,Djama1,Djama2,Djama3}. Indeed,
we have introduced the relativistic quantum Lagrangian written as
\cite {Djama2}
\begin{equation}
L=-m_0c^2 \sqrt{1-f(x){\dot{x}^2 \over c^2}}-V(x)\; ,
\end{equation}
from which, and using the least action principle, we deduce the
expression of the conjugate momentum  \cite {Djama2}
\begin{equation}
\dot{x}{\partial S_0 \over \partial x}=E-V(x)-{m_0^2c^4 \over 
E-V(x)}\; ,
\end{equation}
Then, we derived the Relativistic Quantum Newton's Law
\begin{eqnarray}
\left[(E-V)^2-m_0^2c^4\right]^2+{\dot{x}^2 \over c^2}(E-V)^2
\left[(E-V)^2-m_0^2c^4\right]+{\hbar^2 \over 2} \left[{3 \over
2}\left({\ddot{x} \over \dot{x}}\right)^2-
{\dot{\ddot{x}} \over \dot{x}}\right] \cdot \hskip-10mm&& \nonumber\\
(E-V)^2-{\hbar^2 \over 2}\left(\ddot{x}{dV \over dx}+
\dot{x}^2{d^2V \over dx^2}\right) \left[{(E-V)^2+m_0^2c^4 \over
(E-V)^2-m_0^2c^4}\right] (E-V)^2- {3\hbar^2 \over 4}\cdot
\hskip-1mm&& \nonumber\\
\left(\dot{x}{dV \over dx}\right)^2 \left[{(E-V)^2+m_0^2c^4 \over
(E-V)^2-m_0^2c^4}\right]^2- \hbar^2\left(\dot{x}{dV \over
dx}\right)^2{m_0^2c^4 \over (E-V)^2-m_0^2c^4} =0\; .
\end{eqnarray}
In this paper, we introduce an analogous formalism for a half
spinning particle. In Sec. 2, we present such a formalism, and in
Sec. 3 we conclude and discuss the results.

\vskip\baselineskip
\noindent \textbf{2- Dynamical approach of
the motion\\ of a particle with Spin $1/2$ }
\vskip\baselineskip

First, let us introduce the function $f$ defined in Ref. \cite
{FM1} as

\begin{equation}
f(x)=\left[1-{\hbar^2 \over 2}\left({dh \over
dx}\right)^{-2}\{h(x),x\}\right]^{-1}\; ,
\end{equation}
where $h$ correspond to the reduced action of the particle
($S_{0}$ or $Z_{0}$). Via the function $f$, Eqs. (1) and (2) will
be written as
\begin{eqnarray}
{1 \over 2m_0}\left({dS_0 \over dx }\right)^2{1 \over f_{1}(x)}+
{\hbar^2 \over 2m_{0}}(E-V+m_{0}c^2)^{1 \over 2}\ .
\hskip25mm&& \nonumber\\
\ .{d^2 \over dx^2} \left[(E-V+m_{0}c^2)^{-{1 \over 2}}\right] +
{1 \over 2m_0c^2}\left[m_0^2c^4 -(E-V)^2\right]=0\; ,
\end{eqnarray}
and
\begin{eqnarray}
{1 \over 2m_0}\left({dZ_0 \over dx }\right)^2{1 \over f_{2}(x)}+
{\hbar^2 \over 2m_{0}}(E-V-m_{0}c^2)^{1 \over 2}\ .
\hskip25mm&& \nonumber\\
\ .{d^2 \over dx^2} \left[(E-V-m_{0}c^2)^{-{1 \over 2}}\right] +
{1 \over 2m_0c^2}\left[m_0^2c^4 -(E-V)^2\right]=0\; ,
\end{eqnarray}
which give
\begin{equation}
f_{1}(x)={{c^2 ({dS_{0}/ dx})^2} \over
{(E-V)^2-m_{0}^2c^4-\hbar^2 c^2(E-V+m_{0}c^2)^{1 \over 2}{d^2
\over dx^2}\left[ (E-V+m_{0}c^2)^{-{1 \over 2}}\right]}} \; ,
\end{equation}
and
\begin{equation}
f_{2}(x)={{c^2 ({dZ_{0}/ dx})^2} \over {(E-V)^2-m_{0}^2c^4-\hbar^2
c^2(E-V-m_{0}c^2)^{1 \over 2}{d^2 \over dx^2}\left[
(E-V-m_{0}c^2)^{-{1 \over 2}}\right]}} \; ,
\end{equation}
Now, in order to establish a dynamical approach of the motion, we
introduce the following expressions of the two Lagrangians
$L_{1}$ and $L_{2}$ corresponding to the two RQSHJES$_{1 \over 2}$
\begin{equation}
L_1=-m_0c^2 \sqrt{1-f_{1}(x){\dot{x}^2 \over c^2}}-V(x)\; ,
\end{equation}
\begin{equation}
L_2=-m_0c^2 \sqrt{1-f_{2}(x){\dot{x}^2 \over c^2}}-V(x)\; .
\end{equation}
$L_{1}$ and $L_{2}$ describe the motion of particle in the two
projections of the spin cases ($m_{s}=\pm {1 \over 2}$).

\noindent Using the Least action principle and after integrating,
we get
\begin{equation}
{m_0c^2 \over \sqrt{1-f_{1}(x){\dot{x}^2 \over c^2}}}+V(x)=E\; ,
\end{equation}
and
\begin{equation}
{m_0c^2 \over \sqrt{1-f_{2}(x){\dot{x}^2 \over c^2}}}+V(x)=E\; ,
\end{equation}
where $E$ is an integrating constant representing the total
energy of the particle.

\noindent Remark that, for purely relativistic cases $(\hbar \to
0)$, $f_{1} \to 1$ and $f_{2} \to 1$, Eqs. (17) and (18) reduce
to the well known conservation equation
\begin{equation}
{m_0c^2 \over \sqrt{1-{\dot{x}^2 \over c^2}}}+V(x)=E\; ,
\end{equation}
Now, if we replace the expressions of $f_{1}$ and $f_{2}$ given
by Eqs. (13) and (14) into Eqs. (17) and (18), we find
\begin{eqnarray}
\dot{x}{dS_0 \over dx}=\left(E-V(x)-{m_0^2c^4 \over 
E-V(x)}\right)\ . \hskip50mm\nonumber\\
\ .\sqrt{1-{\hbar^2 c^2 \over
(E-V)^2-m_{0}^2c^4}\sqrt{E-V+m_{0}c^2}{d^2\over dx^2}\left({1
\over \sqrt{E-V+m_{0}c^2}}\right)}\; \; \; ,
\end{eqnarray}
and
\begin{eqnarray}
\dot{x}{dZ_0 \over dx}=\left(E-V(x)-{m_0^2c^4 \over 
E-V(x)}\right)\ . \hskip50mm\nonumber\\
\ .\sqrt{1-{\hbar^2 c^2 \over
(E-V)^2-m_{0}^2c^4}\sqrt{E-V-m_{0}c^2}{d^2\over dx^2}\left({1
\over \sqrt{E-V-m_{0}c^2}}\right)}\; \; \; .
\end{eqnarray}
$dS_0 / dx$ and $dZ_0 / dx$ represent the two conjugate momenta of
the particles with spin ${1 \over 2}$, one particle with spin
projection $m_s=+{1 \over 2}$, the other with $m_s=+{1 \over 2}$.

\noindent Remark that for the constant potentials, the two
momenta reduce to the relativistic quantum momentum (Eq. (8)), so
the degenerating expression disappear for the momenta as well as
for the RQSHJES$_{1 \over 2}$ and the reduced actions \cite
{Djama1}.

\noindent Note that, taking the purely relativistic limit ($\hbar
\to 0$) Eqs. (20) and (21) reduce to the well known conservation
equation (Eq. (19)).

\noindent Remark also that, for the purely quantum cases
$T<<m_0c^2$($T=E-V-m_0c^2$ is the kinetic energy of the
particle), Eqs. (20) and (21) reduce to Eq. (8) given the
expression of the purely quantum conjugate momentum.

Both relativistic quantum Newton's Law can be derived after
replacing both conjugate momenta, given by Eqs. (20) and (21),
into Eqs. (1) and (2). So, we get to the first integral of the
quantum Newton's Law for Spinning particles ($s={1 \over 2 }$)
(FIQNLS$_{1 \over 2}$), equation which is a third order
derivative of coordinate $x$ with respect to time $t$, containing
the constant $E$. Deriving this equation with respect to $x$, we
deduce the QNLS$_{1 \over 2}$ which is a fourth order derivative
of $x$ with respect to $t$. Because the overflowing of the
FIQNLS$_{1 \over 2}$, we do not present it in the present paper.
However, if one want to plot the trajectories of the spinning
particle, he can use the expression of the conjugate momenta
(Eqs. (20) and (21)) grounding himself on the solutions of the
Dirac equation (Eq. (3)). This can be done with the same manner
as it is done in Refs. \cite {B-D2,Djama2} for the particles
without spinning behaviour.

\vskip\baselineskip \noindent \textbf{3- Conclusion }
\vskip\baselineskip

We present, in this paper, a Lagrangian formulation of a
deterministic dynamics of the particle with spin ${1 \over 2}$.
We demonstrate that, as for the no spinning particle case, it is
possible to investigate the dynamical law of motion for the
relativistic quantum phenomena. This is another important step to
build a deterministic approach of quantum mechanics.

\noindent The relativistic quantum conjugate momenta reduce to
the classical one, when $\hbar \to 0$. We demonstrate also that,
each projection of the spin $1 \over 2$ is described by its own
Lagrangian, conjugate momentum, RQSHJE$_{1 \over 2}$ and reduced
action. This means that, for such a particle there is two classes
of trajectories corresponding to the two projections of spin $1
\over 2$.

\noindent For the purely quantum limit $(T<<m_0c^2)$, the
conjugate momenta reduce to two distinguished momenta, which make
the description of the spinning particles with our deterministic
approach always possible. So the spin can be introduced in our
approach even for non relativistic and purely quantum cases. This
point will be more investigated in next works.

Finally, we stress that, for more understanding and completing of
this approach, one must generalize to more than one dimension
problems.

\newpage
\vskip\baselineskip \noindent \textbf{REFERENCES}
\vskip\baselineskip

\end{document}